research paper

# Theoretical and experimental development of a high-conversion-efficiency rectifier at X-band

FEIFEI TAN AND CHANGJUN LIU

Voltage doubler rectifiers are usually applied to systems with high voltage and low current requirement. An X band voltage doubler rectifier has been developed with 72% conversion efficiency. To the best of our knowledge, the obtained rectifying efficiency is the maximum reported to date at X band with Schottky diodes. The working characteristics of the diodes in the voltage doubler rectifier are analyzed in detail. Closed-form equations of diode input impedance and rectifying efficiency are presented and validated using Advanced Design System simulations. The matching network design of the proposed rectifier is based on the closed-form equations. The preliminary rectifying efficiency is predicted by the closed-form equations as well. Measured and simulated results are in good agreement.

## I. INTRODUCTION

Microwave (MW) wireless power transmission technology is based on MW power generation, transmission, and conversion. It has been applied to several applications with advantages, such as solar power satellites, surveillance platforms, wireless sensors, remotely powered vehicles, and radio-frequency identification (RFID) tags. Consequently, wireless power transmission over long distances with high efficiency is required in practice. Rectifiers are a key component of a MW wireless power transmission system. It is located at the receiving end whose MW–DC conversion efficiency is critical to the overall system efficiency.

There are numerous published reports on research involving rectifiers [1–3]. With respect to the operating frequency band, they work at single frequency [4, 5], multi-frequency [6–9], or broadband [10]. With respect to the topology structure, they work with single diode, dual diodes [11–14], or diode arrays [15].

Current studies show that the optimal MW–DC conversion efficiency of a rectifier is related to the operating frequency band. In 1977, an aluminum bar-type rectenna made by Brown achieved more than 90% efficiency at S band [1]. Recent studies show that efficiency can reach up to 75% at C band [16, 17]. A rectifier with single series diode achieves an efficiency of 71.9% at X band [18], while it is 53 and 60% at Ka band in [19, 20], respectively. The plotting of efficiency versus operating frequency of the state-of-the-art in the field of rectifiers is shown in Fig. 1. From Fig. 1, the highest conversion efficiencies are not the same at different frequencies. The general trend is that, as the operating frequency increases, the highest MW–DC conversion efficiency decreases. It is meaningful to make comparisons between rectifiers at the same operating frequency band.

In this paper, we present a novel voltage doubler rectifier operating at X band for a MW power transmission system. A detailed comparison between the proposed rectifier and relevant rectifiers is presented in Table 1. From Table 1, the majority of previous studies used Schottky diode as a rectifying device. The conversion efficiencies of diode-based rectifiers are usually higher than those of transistor- and Monolithic Microwave Integrated Circuit (MMIC)-based rectifiers. However, their input powers are lower. Among these diode-based rectifiers, the proposed rectifier not only achieves higher conversion efficiency, but also works at relatively higher power level, making it distinct from the previous work.

A theoretical analysis of diodes in voltage doubler rectifiers is presented based on the theory in [25]. This model comprised a single diode and a parallel load resistance. The diode's characteristics were described by its SPICE parameters and the incident sine waveform to the rectifier. Valenta et al. [30] also modeled energy-conversion efficiency circuits using the one-diode model. It considered the conversion efficiency under high peak-to-average power ratio (PAPR) signals instead of a continuous waveform excitation. The duty cycle of a signal with intermittent transmission was taken into account in the derivation for the closed-form equations.

However, the models in references [25, 30] are not suitable for the analysis of a voltage doubler rectifier, which has two diodes, and the waveforms across them are different in the

School of Electronics and Information Engineering, Sichuan University, Chengdu 610064, China. Phone: +86-28-85463882
**Corresponding author:**
C. Liu
Email: cjliu@ieee.org

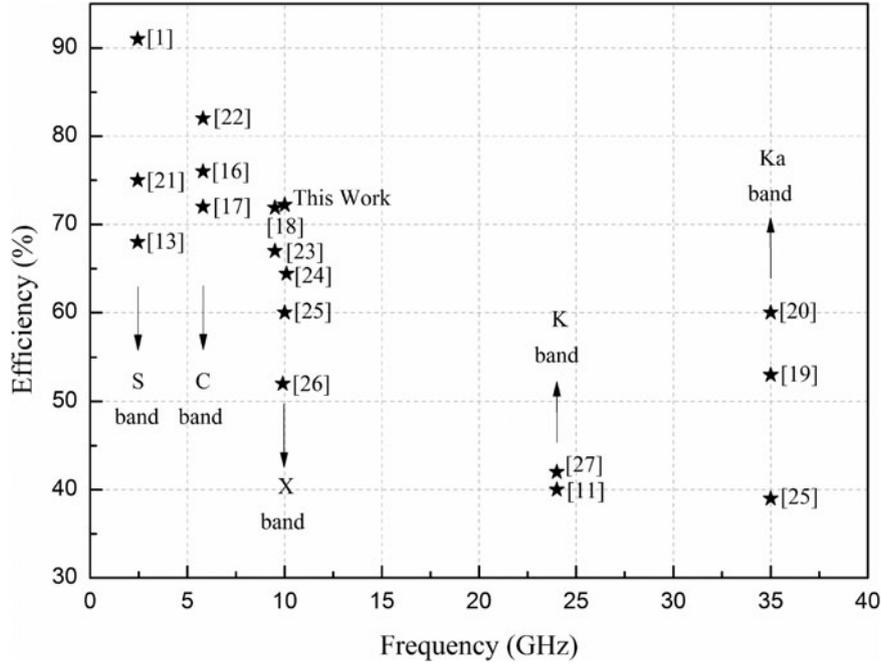

**Fig. 1.** Efficiency with respect to frequency of the state-of-the-art in the field of rectifiers.

single diode condition. This paper will expand the one-diode model analysis further to voltage doubler model analysis. The excitation of a rectifier in a MW power transmission system is generally continuous waveform. Therefore, we use this excitation in the analysis rather than a high PAPR signal to meet the application in a MW power transmission system. The advantage of the analysis presented in this paper is its validity for a MW voltage doubler rectifier design. It offers a useful tool for rectifying efficiency prediction and circuit design.

## II. THEORETICAL ANALYSIS

A typical block diagram of a rectifier is shown in Fig. 2. A low pass filter in front of the diode ensures that the fundamental frequency MW goes through and higher-order harmonics generated from the nonlinear diode are rejected back to the rectifying circuit. It could be eliminated due to the trade-off between the recycled power and the additional higher insertion loss in some designs [13]. A DC pass filter is applied to reject the harmonics and make the DC output voltage smooth. A matching network is used for transforming the circuit input impedance to be matched to the source. The design of impedance matching network is mainly focused on the diode, which has dynamic input impedance due to the nonlinearity.

In the theoretical analysis presented in this paper, the filters and matching network are considered to be perfectly designed. The harmonics produced by the diodes during rectification are assumed to be recycled completely. There are no harmonics energy losses and insertion loss in the rectifying circuit. Consequently, the lumped-parameter model of a voltage doubler rectifier may be considered and analyzed without the input impedance matching network/filter, as shown in Fig. 3. The analysis of the characteristics of the diodes in this kind of circuit is as follows.

### A) Diode conduction angle

Figure 4 shows the MW voltage-current waveforms operating across the diode and its junction. The incident MW sine voltage is $V_p \sin \omega t$. The voltage across the diode is shown as

$$V_I = -V_T + V_P \sin \omega t, \quad (1)$$

where $V_T$ is the DC bias voltage across the diode, which is the combination of DC output voltage $V_D$ and diode conduction voltage $V_F$. Here, $V_T = V_D + V_F$.

The diodes in Fig. 3 are expressed by their equivalent circuit parameters as shown in Fig. 5. This model comprises a series resistance $R_s$, a diode junction resistance $R_j$, and a junction capacitance $C_j$. $R_L$ is the load resistance. Power losses occur on both the series resistance and diode junction. Assuming that $R_j$ is close to zero in the conducting state and infinite in the cutoff state, the loss through diode junction has been neglected during the cutoff period because the current is zero.

**Table 1.** Comparison table of the proposed rectifier and related work at X band.

| Time | Reference | Device | Power | Efficiency (%) |
|------|-----------|--------|-------|----------------|
| 1992 | [25] | DMK6606, Alpha industries | 20.79 dBm | 60 |
| 2011 | [28] | HSMS-8202 Avago | 245 uW/cm² | 21 |
| 2012 | [29] | Diode-connected NMOS transistors | −8 dBm | 3.1 |
| 2014 | [24] | GaN MMICs fabricated | 34.14 dBm | 64.4 |
| 2014 | [23] | HSMS-8101 Avago | 15.5 dBm | 67 |
| 2014 | [18] | HSMS-8101 Avago | 17 dBm | 71.9 |
| 2015 | [26] | MMIC | >8 W | 52 |
| 2016 | This work | HSMS-286C Avago | 19.4 dBm | 72.2 |

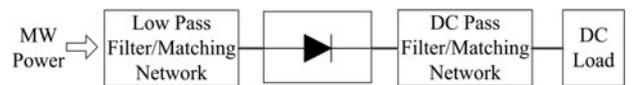

**Fig. 2.** Typical block diagram of a rectifier.

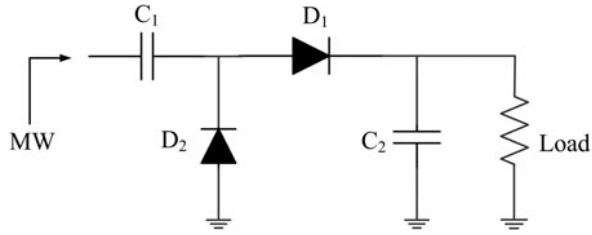

Fig. 3. Lumped-parameter model of a voltage doubler rectifier.

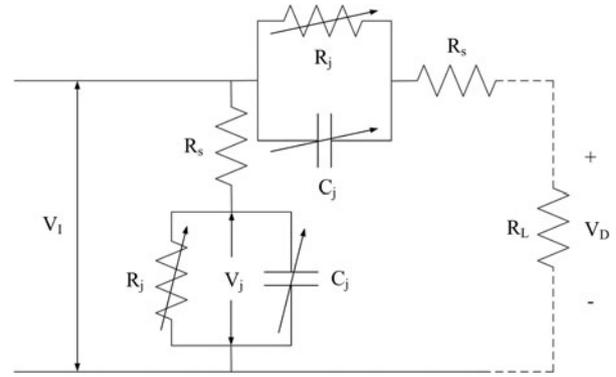

Fig. 5. Equivalent circuit model of the voltage doubler rectifier.

Due to the diode junction, $V_j$ lags behind $V_I$ by a slight order of $\phi$ as shown in Fig. 4, $V_j$ can be expressed as in equation (2). $V_{jo}$ and $V_{j1}$ are the DC and MW parts of the input voltage operating on the diode junction, respectively.

$$V_j = \begin{cases} -V_{jo} + V_{j1}\sin(\omega t - \phi), & \text{diode off} \\ V_F, & \text{diode on} \end{cases}, \quad (2)$$

$\theta_i$ is the diode conduction angle. In one cycle, the diode is in the conducting state when $\theta \in [(\pi - \theta_i)/2, (\pi + \theta_i)/2]$ and in the cutoff state during the rest of the period. Applying Kirchhoff's law on the DC cycle in Fig. 5, we have equation (3).

$$V_F + 2I_D R_s + V_D + \bar{V}_j = 0, \quad (3)$$

where $I_D = V_D/R_L$. Then, we obtain:

$$V_D = -\frac{V_F + \bar{V}_j}{1 + (2R_s/R_L)}. \quad (4)$$

The DC output voltage is determined by diode forward conduction voltage $V_F$ and diode average junction voltage over one period:

$$\bar{V}_j = \frac{1}{2\pi}\left(\int_{\pi-\theta_i/2}^{\pi+\theta_i/2} V_F d\theta + \int_{\pi+\theta_i/2}^{5\pi-\theta_i/2} (-V_{jo} + V_{j1}\sin\theta)d\theta\right). \quad (5)$$

The first item and the second item in the brackets represent the voltage integration during conduction and cutoff period, respectively. From (5), we achieve:

$$\bar{V}_j = \frac{\theta_i}{2\pi}V_F + \frac{\theta_i - 2\pi}{2\pi}V_{jo} - \frac{\sin(\theta_i/2)}{\pi}V_{j1}, \quad (6)$$

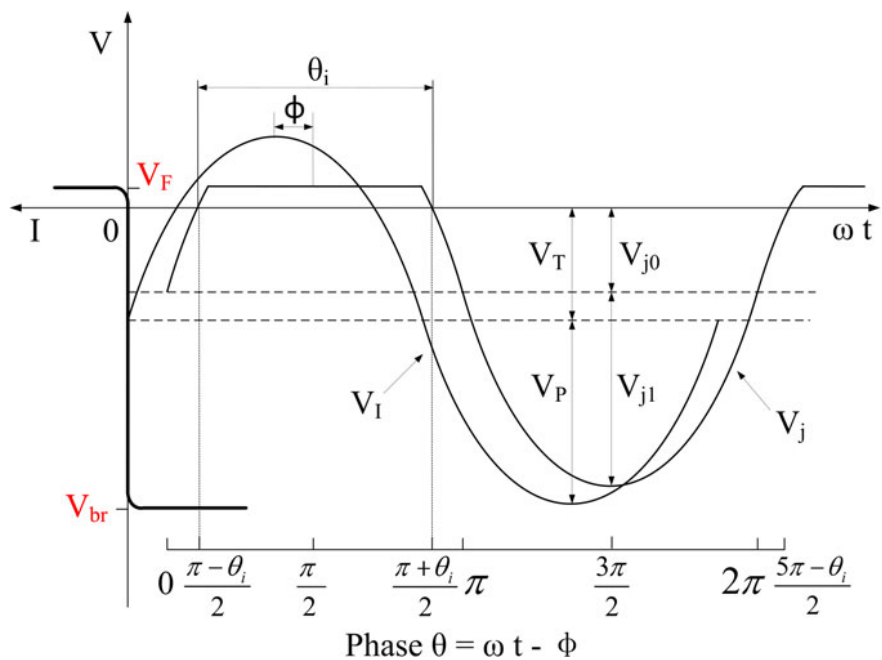

Fig. 4. Waveforms of voltage on diode and its junction, and curve of diode current-voltage characteristic.

$C_j$ is charged when diode is cutoff. The charging current is

$$I_c = \frac{dC_j V_j}{dt}. \quad (7)$$

Expand $C_j$ by Fourier series,

$$C_j = C_0 + C_1 \sin(\omega t - \phi) + C_2 \sin(2\omega t - 2\phi) + \cdots. \quad (8)$$

The higher-order harmonics are neglected according to the assumptions. Substitute (8) into (7),

$$I_c = \omega(C_0 V_{j1} - V_{j0} C_1) \cos(\omega t - \phi). \quad (9)$$

When the diode is cutoff, we obtain:

$$V_I = V_j - I_c R_s. \quad (10)$$

Equations (9) and (10) yield

$$\begin{cases} V_{j0} = V_F + V_D \\ V_{j1} = V_P \cos \phi \end{cases}, \quad (11)$$

where $\phi$ is the phase difference between $V_I$ and $V_j$, which is close to zero. When the diode switches from cutoff to on, $V_j = V_F$. Then,

$$-V_{j0} + V_{j1} \sin\left(\frac{\pi}{2} + \frac{\theta_i}{2}\right) = V_F. \quad (12)$$

Substituting $V_{j0}$ and $V_{j1}$ of (11) into (12), $V_P$ is expressed as:

$$V_P = \frac{2V_F + V_D}{\cos(\theta_i/2)}. \quad (13)$$

$\bar{V}_j$ and $V_D$ are expressed by the given variables from (4), (6), (11), and (12). The conduction angle of the diode satisfies:

$$\tan\frac{\theta_i}{2} - \frac{\theta_i}{2} = \frac{2\pi R_s}{R_L(1 + (2V_F/V_D))}. \quad (14)$$

This is a transcendental equation to obtain the conduction angle. It is determined by diode forward conduction voltage $V_F$, series resistance $R_s$, output DC voltage $V_D$, and load resistance $R_L$. $R_s$ and $V_F$ are mainly determined by the diode characteristics, while $V_D$ and $R_L$ are mainly determined by the input power level.

## B) Diode impedance

The diode dynamic input impedance is

$$Z(\omega t) = \frac{V_P}{I(\omega t)}, \quad (15)$$

$$I(\omega t) = |I| \sin \omega t - j|I| \cos \omega t. \quad (16)$$

The second item in (16) occurs only when the diode is in the cutoff state. The average current over one period is

$$I = \frac{1}{\pi}\left\{\int_{\pi-\theta_i/2}^{\pi+\theta_i/2} \left|\frac{V_I - V_F}{R_s}\right| \sin \omega t\, d\theta \right. \\ \left. + j\int_{\pi+\theta_i/2}^{5\pi-\theta_i/2} \left|\frac{V_I - V_F}{R_s}\right| \cos \omega t\, d\theta\right\}. \quad (17)$$

For the first item in the brackets, $V_I$ will be substituted based on equation (1). The second item $|V_I - V_F/R_s|$, will be replaced by the charging current $I_c$. According to equations (2), (7), and (11), $I_c$ is

$$I_c = \omega C_j V_P \cos(\omega t - \phi). \quad (18)$$

Substituting $V_I$ and $I_c$ into (17),

$$I = \frac{2V_F + V_D}{2\pi R_s}\left(-4\sin\frac{\theta_i}{2} + \frac{\theta_i + \sin\theta_i}{\cos(\theta_i/2)}\right) \\ + j\frac{\omega C_j (2V_F + V_D)}{2\pi \cos(\theta_i/2)}\left(\pi - \frac{\theta_i}{2} + \frac{\sin\theta_i}{2}\right). \quad (19)$$

Then, the diode input impedance is as follows:

$$Z = \frac{2\pi R_s}{\theta_i - \sin\theta_i + j2R_s \omega C_j(\pi - (\theta_i/2) + \sin\theta_i/2)}. \quad (20)$$

We assume that the circuit is well matched at input power and the optimal DC output voltage is 0.9 $V_{br}$. The diode input impedance and conduction angle were calculated using (14) and (20). The results are shown in Fig. 6, and the diode parameters are listed in Table 2. As shown in Fig. 6, the diode conduction angle increases along with the input power. The resistance plot has a vertex around 30 mW, which is considered to be caused by the diode threshold voltage. When the input power is low, the diode is mostly off and presents relatively high resistance. As the input power increases, the effect of threshold voltage and the diode resistance decreases. In either case, the diode presents capacitive reactance.

To validate the closed-form equation, the calculated results are compared with the results simulated by the ADS software. The simulation uses ideal components without package parasitic effects, and the Large Signal S Parameter (LSSP) simulation engine is applied to obtain the input impedances. The resistance values calculated with the closed-form equation are close to the simulated ones as shown in Fig. 6. The reactance difference between simulated and calculated results is considered to be caused by the difference between junction capacitance models, which substantially affect the diode reactance. The closed-form equation is accurate enough to build the preliminary rectifying circuit topology. It is important to provide a guideline for MW nonlinear rectifier designing rather than the circuit design based on random optimization, which usually takes a much longer time.

Table 2. Parameters of HSMS 286 Schottky diode and circuit capacitors.

| Diode | | | | | Capacitance | |
|---|---|---|---|---|---|---|
| $V_{br}$ (volts) | $R_s$ ($\Omega$) | $C_{j0}$ (pF) | $I_s$ (pAmps) | $V_F$ (V) | $C_1$ (pF) | $C_2$ (pF) |
| 7 | 7 | 0.18 | 5E-8 | 0.35 | 5.6 | 20 |

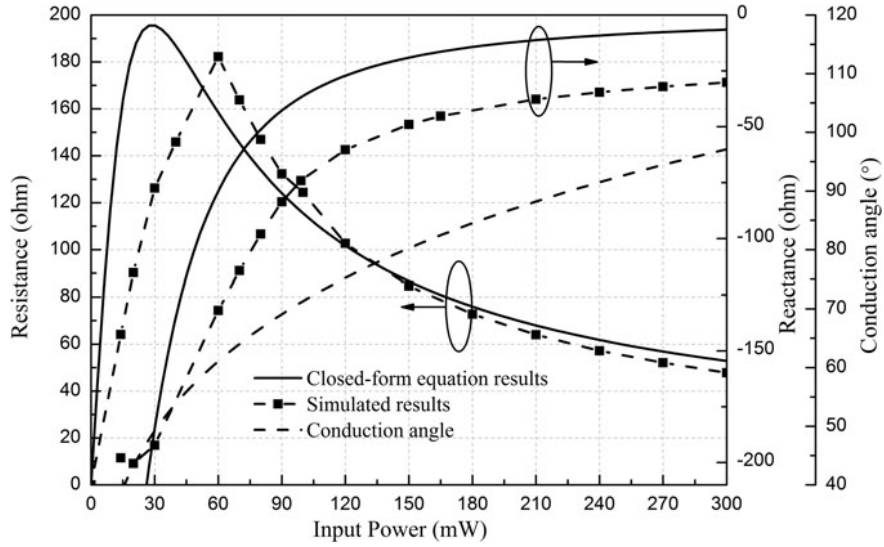

Fig. 6. Comparison of diode impedances between calculation using the closed-form equation and ADS simulation, and calculated conduction angles.

### C) Diode efficiency

Power losses of a MW rectifier come from three parts: dielectric and conductor losses, radiation loss, and diode loss, in which diode loss usually plays a major role. The diode efficiency is defined as

$$\eta = \frac{P_c}{P_c + L_{R_s} + L_j}, \qquad (21)$$

where $P_c$ is the DC output power, and $L_{Rs}$ and $L_j$ are the power losses on the diode series resistance and diode junction, respectively. According to Fig. 5, the DC output power, diode series resistance loss, and diode junction loss are defined as follows:

$$P_c = \frac{V_c^2}{R_L}, \qquad (22)$$

$$L_{R_s} = \frac{1}{\pi}\left\{\int_{\pi-\theta_i/2}^{\pi+\theta_i/2} \frac{(V_I - V_F)^2}{R_s} d\theta + \int_{\pi+\theta_i/2}^{5\pi-\theta_i/2} \frac{(V_I - V_j)^2}{R_s} d\theta\right\}, \qquad (23)$$

$$L_j = \frac{1}{\pi}\int_{\pi-\theta_i/2}^{\pi+\theta_i/2} V_F \frac{(V_I - V_F)}{R_s} d\theta. \qquad (24)$$

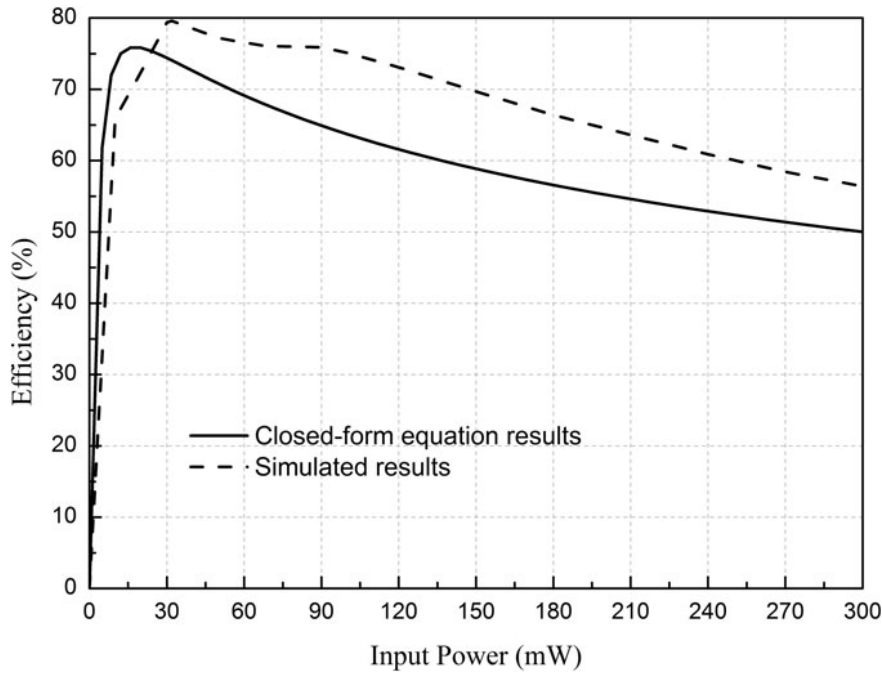

Fig. 7. Comparison of diode conversion efficiency calculated from the closed-form equation and simulated from the ADS.

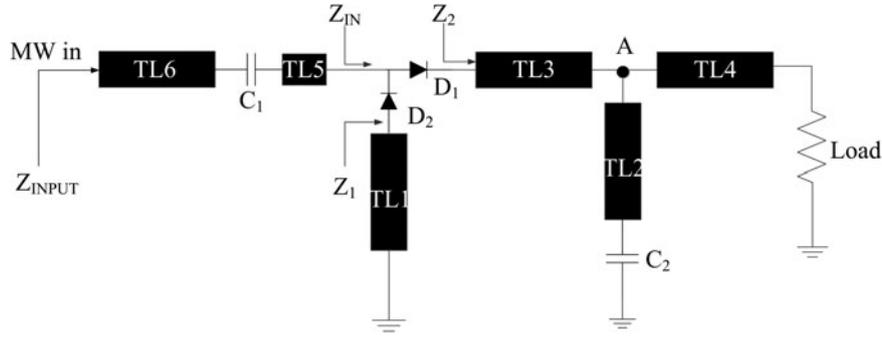

**Fig. 8.** The topology of the designed X band voltage doubler rectifier.

By substituting (22)–(24) into (21), we obtain the diode efficiency:

$$\eta = \frac{V_c^2}{\begin{array}{l}(2R_L(V_F + V_c)^2/\pi R_s)(2\theta_i - 8\tan(\theta_i/2) + \\ ((\theta_i + \sin\theta_i)/\cos^2(\theta_i/2))) + (4\omega^2 C_j^2 R_s R_L \\ (V_F + V_c)^2/\pi \cos^2(\theta_i/2))(\pi - (\theta_i/2) + \\ (\sin\theta_i/2)) + (2V_F R_L(V_F + V_c)/\pi R_s) \\ (2\tan(\theta_i/2) - \theta_i) + V_c^2\end{array}}. \quad (25)$$

To validate the closed-form equation on rectifying efficiency, the calculated results are compared with the simulated ones from the ADS software. Each circuit designed in the ADS was optimized to its maximum rectifying efficiency at the specified input MW power. The simulation uses ideal components without package parasitic effects and the Harmonic Balance (HB) simulation engine to obtain the rectifying efficiency. As shown in Fig. 7, the calculated rectifying efficiency agrees well with simulated results at 10 GHz. The maximum error in efficiency is 10%. The maximum rectifying efficiencies are close to each other, while there is a shift on the input MW power. However, the closed-form equation predicts the general efficiency of a voltage doubler rectifier with the specified diode at an arbitrary frequency. It is meaningful since it can be applied to diode selection and estimating the efficiency level before accurate designing.

**Table 3.** The supposed values of $TL_1$–$TL_3$

| Parameter | Length | $Z_{in}$ |
|---|---|---|
| $TL_1$ | $l_1 + n\lambda/2$ | $j26.9$ |
| $TL_2$ | $n\lambda/2$ | $0$ |
| $TL_3$ | $l_2 + n\lambda/2$ | $j29.0$ |

### III. RECTIFIER DESIGN

$$Z_{in} = Z_o \frac{Z_L + jZ_o \tan\theta}{Z_o + jZ_L \tan\theta}. \quad (26)$$

The topology of the proposed rectifier is shown in Fig. 8. The output DC filter is constituted by TL2 and $C_2$. Capacitor $C_2$ is regarded as a short circuit at MW frequency. The length of TL2 is half-wavelength at the fundamental frequency so as to present a short-circuit point "A" at this frequency and its harmonics. MW power is reflected back to the diode at point A. The relevant equation is shown in (26), in which $Z_0$ is the characteristic impedance of a transmission line, which is associated with its width; $Z_L$ is the load impedance of the line; $\theta$ is the electrical length of the line. Therefore, whatever $Z_0$ is, the $Z_{in}$ of TL2 is fixed at zero in this case. $Z_0$ is set to be 50 Ω in the preliminary design for TL2.

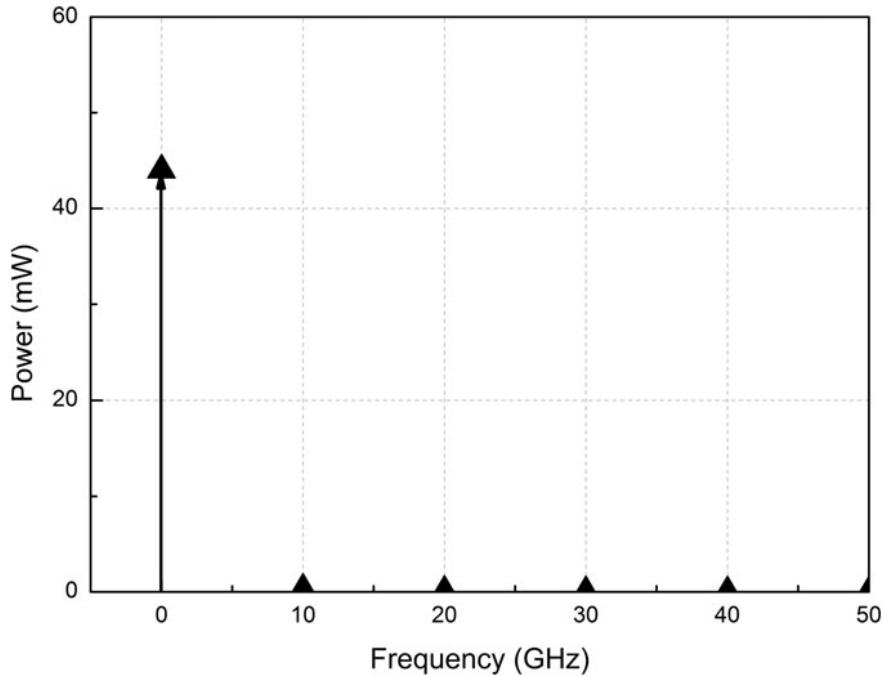

**Fig. 9.** DC and harmonic power at the output port of the designed circuit.

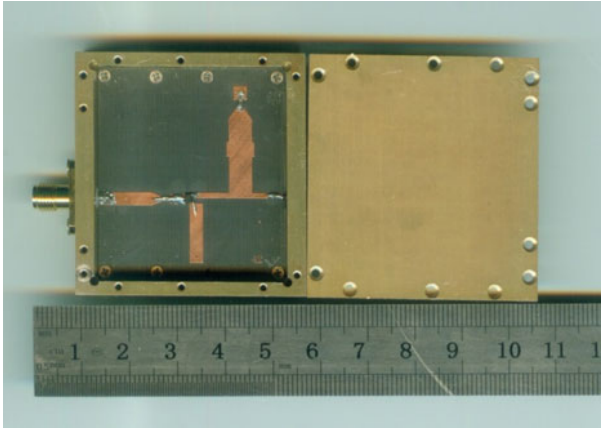

Fig. 10. The fabricated voltage doubler rectifier with a cavity.

TL4 is designed for conducting DC current. It is shorted by point A at MW frequency. Thus, its width and length has no contribution to the circuit impedance matching design at fundamental frequency.

TL3 is a segment of transmission line with a short-circuit ending (point A). Its characteristic impedance is set to be equal to TL4 to avoid a steep variation in microstrip line widths, which leads to power losses due to the electromagnetic radiation and reflection. Its length is calculated from equation (26). In this design, the equivalent input impedance of TL3 is set to be the conjugate of diode reactance. In Fig. 6, the diode impedance is $100.6 - j27.2$ Ω at an input MW power of 120 mW. If the input impedance of TL3 is $j27.2$ Ω, the reactance of diode $D_1$ is compensated and $Z_2$ in Fig. 8 has to be a pure resistance. TL1 is applied to compensate for the reactance of diode $D_2$ as well. Consequently, $Z_{IN}$ is pure resistance.

$Z_{INPUT}$ is the target value of the impedance transformation. It is usually a pure resistance, for instance, 50 Ω. A quarter-wavelength microstrip line can be added to transform $Z_{IN}$ to $Z_{INPUT}$ by tuning its width directly if $Z_{IN}$ is pure resistance as well. Therefore, the advantage offered by the above procedures is to make it easier for the impedance matching design.

In this work, $Z_{IN}$ in Fig. 8 is supposed to be 50.3 Ω, which is matched with the signal source. Hence, it is no need to add any lines for impedance matching design in front of the diodes. Therefore, TL1 and TL3 constitute the impedance compensation network already.

The components in front of the diodes have no contribution to the impedance matching. TL5 was used for connecting $C_1$ and the diodes. Its length is less than one-tenth of the wavelength and can be seen as a lumped element, which is equivalent to a short circuit. Its characteristic impedance is set to be 50 Ω in the preliminary design. The reactance of $C_1$ is much less than 50 Ω and is neglected. TL6 is a microstrip line with a characteristic impedance of 50 Ω, which is used for connecting the source. Its length will not affect anything according to equation (26).

Agilent ADS simulations were performed to optimize the voltage doubler rectifier further. The finalized input impedances of TL1 and TL3 are $7.6 + j26.9$ Ω and $9.5 + j29.0$ Ω, respectively. This result coincides with the theoretical results (Table 3).

The harmonic power was monitored at the output port by probes in the ADS simulation. In Fig. 9, except for DC power,

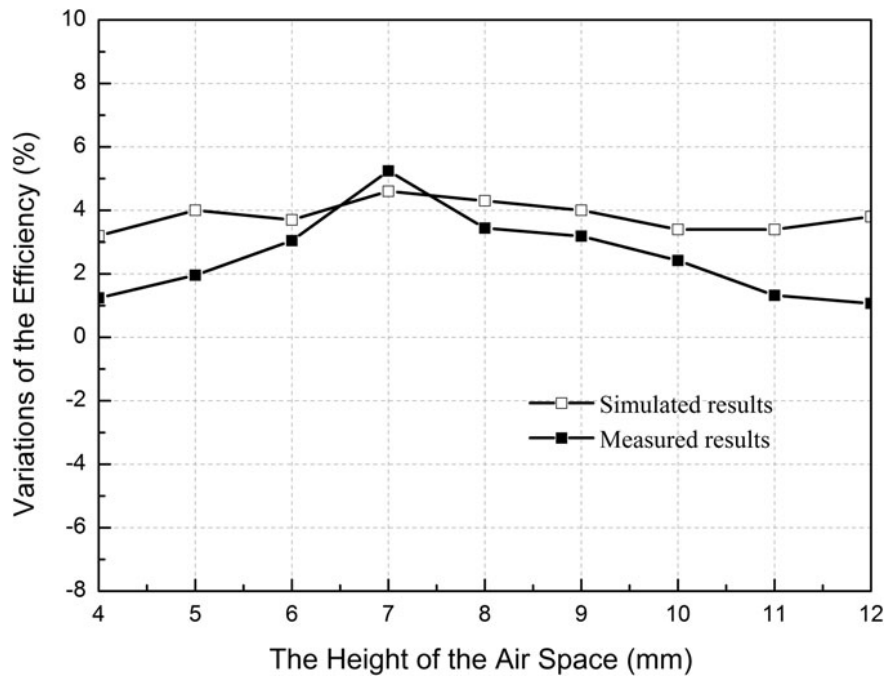

Fig. 11. The simulated and measured results of efficiency varying with air layer thickness ranging from 4 to 12 mm.

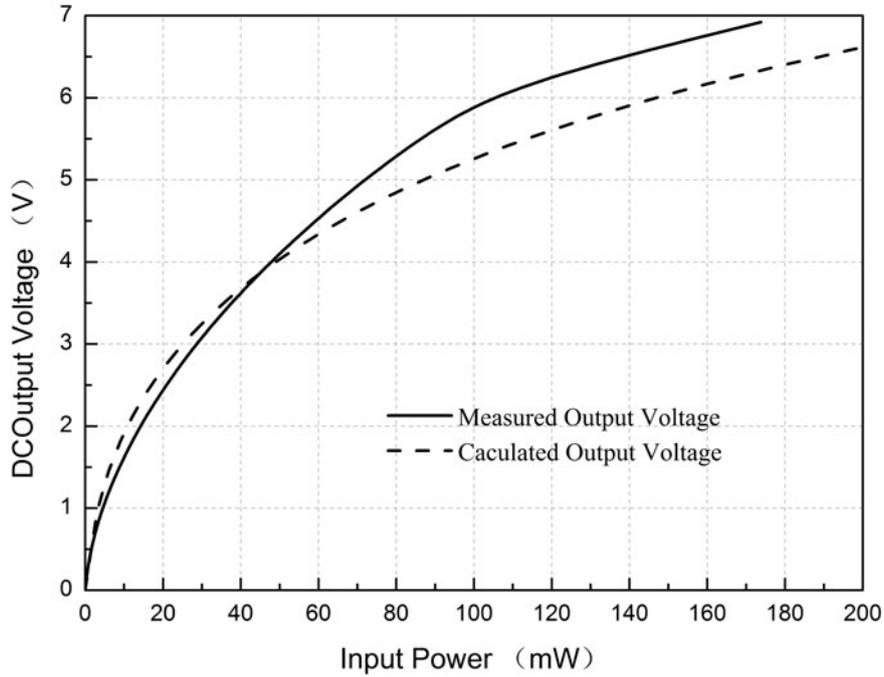

**Fig. 12.** The measured and calculated output DC voltage of the proposed rectifier with respect to the input power.

power at fundamental and harmonic frequencies is close to zero. The rectifier is able to significantly suppress the harmonics produced during rectifying.

The voltage doubler rectifier is fabricated on a 1-mm F4B-2 substrate with relative dielectric constant 2.65 and loss tangent 0.005. Then, AVAGO HSMS-286C Schottky diodes and ATC 600S series capacitors are applied. Their parameters and capacitances are listed in Table 2. The shortest length of TL1 and TL3 is around 3 mm, which has been lengthened with a half-wavelength to avoid possible coupling between microstrip lines. Another advantage is to enhance the heat dissipation of the diode so as to improve the stability. For rectifiers at high power system, heat dissipation is necessary.

Because of the operating MW frequency, we took circuit radiation into consideration. Therefore, a metal cavity was applied to recycle the radiated power. It is similar to an antenna reflecting plate design. The height of the air layer determines the cavity's ability in terms of power recovery. The designed rectifier and its cavity are shown in Fig. 10. We introduced perfect conductive surfaces around the circuit in the ADS software to optimize the height of the air layer in the cavity. The simulated and measured results of the rectifying efficiency variation are shown in Fig. 11. Each circuit was optimized to its maximum efficiency by tuning the input power and DC load resistance at a specific height of the air layer. When the height of the air layer is 7 mm, the highest efficiency is achieved. A rectifier with a metal cavity offers several advantages, such as it prevents electrostatic breakdown, it is dust-proof, and it is moisture-proof.

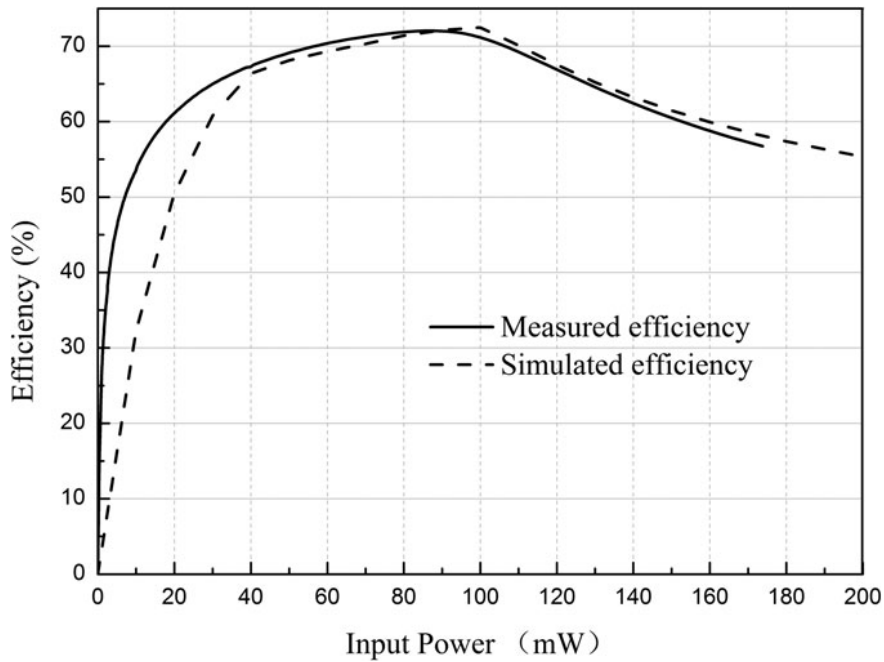

**Fig. 13.** Rectifier conversion efficiency by measurement and using the simulation software ADS. The load resistance is 486 Ω.

The fabricated circuit was measured with an Agilent E8267C MW vector signal generator. For comparative analysis, the measured DC voltages are compared with the ones calculated by the closed-form equation, as shown in Fig. 12. In this calculation, we use the efficiencies calculated in Fig. 7, and fix the load resistance on 486 Ω, which is equal to the measured optimal load resistance. The two plots have a similar trend.

The measured efficiencies are shown in Fig. 13. The proposed circuit achieves its highest efficiency of 72% at 87 mW with a DC load resistance of 486 Ω. The ADS simulation was performed to obtain the efficiency of the rectifier in the cavity. Measured results are in good agreement with simulated ones.

A conclusion is drawn according to the comparison of calculation, simulation and measurement. The efficiency closed-form equation is applied to predicting the optimal efficiency of a voltage doubler rectifier. Another advantage is to make diode selection based on the calculated optimal rectifying efficiency without long-time optimization in the ADS. The closed-form equation on impedance is helpful to build the preliminary circuit topology. The following optimization is quick due to the almost accurate preliminary circuit topology.

## IV. CONCLUSION

A rectifier has been developed with a measured conversion efficiency of 72% at 10 GHz at an input power of 87 mW. Closed-form equations of diode impedance and conversion efficiency of a MW voltage doubler rectifier have been presented. Simulations in ADS have been performed to validate the closed-form equation. The impedance matching network of the proposed rectifier was built preliminarily based on the closed-form equation. A metal cavity has been added for recycling radiated MW power. The highest measured conversion efficiency coincides with the calculated one. The presented rectifier will be applied to a practical MW power transmission system. Moreover, an impedance compression network [31] may be combined with the proposed rectifier to be applied in a higher MW power transmission system.

## ACKNOWLEDGEMENT

This work was supported in part by the China 973 program under Grant 2013CB328902, and the NSFC under Grant 61271074.

## REFERENCES

[1] Brown, W.C.: Chapter 2.2.4, Electronic and Mechanical Improvement of the Receiving Terminal of a Free-space Microwave Power Transmission System, NASA Sti/recon Technical Report N, 40, 1977.

[2] Matsunaga, T.; Nishiyama, E.; Toyoda, I.: 5.8 GHz stacked differential rectenna suitable for large-scale rectenna arrays with DC connection. IEEE Trans. Antennas Propag., **63** (2015), 5944–5949.

[3] Chou, J.-H.; Lin, D.-B.; Weng, K.-L.; Li, H.-J.: All polarization receiving rectenna with harmonic rejection property for wireless power transmission. IEEE Trans. Antennas Propag., **62** (2014), 5242–5249.

[4] Lorenz, C.H.P. et al.: Breaking the efficiency barrier for ambient microwave power harvesting with heterojunction backward tunnel diodes. IEEE Trans. Microw. Theory Tech., **63** (2015), 4544–4555.

[5] Liu, C.; Guo, Y.-X.; Sun, H.; Xiao, S.: Design and safety considerations of an implantable rectenna for far-field wireless power transfer. IEEE Trans. Antennas Propag., **62** (2014), 5798–5806.

[6] Jiang, W.; Liu, C.; Yu, C.; Tan, F.: A novel dual-frequency microwave rectifier at 2.45 and 5.8 GHz with harmonic recycling. J. Electromagn. Waves Appl., **27** (2013), 707–715.

[7] Lu, P.; Yang, X.-S.; Li, J.-L.; Wang, B.-Z.: A compact frequency reconfigurable rectenna for 5.2- and 5.8-GHz wireless power transmission. IEEE Trans. Power Electron., **30** (2015), 6006–6010.

[8] Ren, Y.-J.; Farooqui, M.F.; Chang, K.: A compact dual-frequency rectifying antenna with high-orders harmonic rejection. IEEE Trans. Antennas Propag., **55** (2007), 2110–2113.

[9] Kuhn, V.; Lahuec, C.; Seguin, F.; Person, C.: A multi-band stacked RFenergy harvester with RF-to-DC efficiency up to 84%. IEEE Trans. Microw. Theory Tech., **63** (2015), 1768–1778.

[10] Song, C.; Huang, Y.; Zhou, J.; Zhang, J.; Yuan, S.; Carter, P.: A high-efficiency broadband rectenna for ambient wireless energy harvesting. IEEE Trans. Antennas Propag., **63** (2015), 3486–3495.

[11] Ladan, S.; Hemour, S.; W u, K.: Towards millimeter-wave high-efficiency rectification for wireless energy harvesting, in IEEE Int. Wireless Symp. (IWS), Beijing, 2013.

[12] Khonsari, Z.; Björninen, T.; Tentzeris, M.M.; Sydanheimo, L.; Ukkonen, L.: 2.4 GHz inkjet-printed RF energy harvester on bulk cardboard substrate, in Radio and Wireless Symp. (RWS), San Diego, CA, 2015.

[13] Olgun, U.; Chen, C.-C.; Volakis, J.L.: Investigation of rectenna array configurations for enhanced RF power harvesting. IEEE Antennas Wirel. Propag. Lett., **10** (2011), 262–265.

[14] Heikkinen, J.; Kivikoski, M.: A novel dual-frequency circularly polarized rectenna. IEEE Antennas Wirel. Propag. Lett., **2** (2003), 330–333.

[15] Zhang, B.; Zhao, X.; Yu, C.; Huang, K.; Liu, C.: A power enhanced high efficiency 2.45 GHz rectifier based on diode array. J. Electromagn. Waves Appl., **25** (2011), 765–774.

[16] Ren, Y.-J.; Chang, K.: 5.8-GHz circularly polarized dual-diode rectenna and rectenna array for microwave power transmission. IEEE Trans. Microw. Theory Tech., **54** (2006), 1495–1502.

[17] Yu, C.; Tan, F.; Liu, C.: A C-band microwave rectenna using aperture-coupled antenna array and novel Class-F rectifier with cavity. J. Electromagn. Waves Appl., **29** (2015), 977–991.

[18] Shin, J.; Seo, M.; Choi, J.; So, J.; Cheon, C.: A compact and wideband circularly polarized rectenna with high efficiency at X-band. Prog. Electromagn. Res., **145** (2014), 163–173.

[19] Chiou, H.-K.; Chen, I.-S.: High-efficiency dual-band on-chip rectenna for 35- and 94-GHz wireless power transmission in 0.13-m CMOS technology. IEEE Trans. Microw. Theory Tech., **58** (2010), 3598–3606.

[20] Li, Z.; Wen, G.: Time-domain analysis of Ka-band rectenna. J. Microw., **2** (1998), 134–141.

[21] Douyère, A.; Luk, J.-D.L.S.; Alicalapa, F.: High efficiency microwave rectenna circuit: modelling and design. Electron. Lett., **44** (2008), 1409–1410.

[22] McSpadden, J.O.; Fan, L.; Chang, K.: Design and experiments of a high-conversion-efficiency 5.8-GHz rectenna. IEEE Trans. Microw. Theory Tech., **46** (1998), 2053–2060.

[23] Kim, Y.; Yoon, Y.J.; Shin, J.; So, J.: X-band printed rectenna design and experiment for wireless power transfer, in 2014 IEEE Wireless Power Transfer Conf. (WPTC), Jeju, South Korea, 2014.

[24] Litchfield, M.; Schafer, S.; Reveyrand, T.; Popovic, Z.: High-efficiency X-band MMIC GaN power amplifiers operating as rectifiers, in 2014 IEEE MTT-S Int. Microwave Symp., Tampa, FL, 2014.

[25] Yoo, T.-W., Changk, K.: Theoretical and experimental development of 10 and 35 GHz rectennas. IEEE Trans. Microw. Theory Tech., **40** (1992), 1259–1266.

[26] Schafer, S.; Coffey, M.; Popovic, Z.: X-band wireless power transfer with two-stage high-efficiency GaN PA/rectifier, in 2015 IEEE Wireless Power Transfer Conf., Boulder, CO, 2015.

[27] Ladan, S.; Guntupalli, A.B.; Wu, K.: A high efficiency 24 GHz rectenna development towards millimeter-wave energy harvesting and wireless power transmission. IEEE Trans. Circuits Syst. I: Regular Papers, **61** (2014), 3358–3366.

[28] Monti, G.; Tarricone, L.; Spartano, M.: X-band planar rectenna. IEEE Antennas Wirel. Propag. Lett., **10** (2011), 1116–1119.

[29] Senadeera, P.M. et al.: X-band energy harvester with miniaturized on-chip slot antenna implemented in 0.18- um RF CMOS, in 2012 IEEE Int. Conf. on Ultra-Wideband, Syracuse, NY, 2012.

[30] Valenta, C.R.; Morys, M.M.; Durgin, G.D.: Theoretical energy-conversion efficiency for energy-harvesting circuits under power-optimized waveform excitation. IEEE Trans. Microw. Theory Tech., **63** (2015), 1758–1767.

[31] Xu, J.; Ricketts, D.S.: An efficient, watt-level microwave rectifier using an impedance compression network (ICN) with applications in outphasing energy recovery systems. IEEE Microw. Wirel. Compon. Lett., **23** (2013), 542–544.

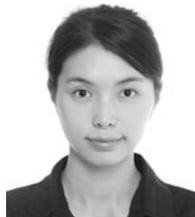

**Feifei Tan** earned her BS in Electronics Information Engineering from Sichuan University of China in 2011. She is currently working toward her Ph.D. in Communication and Information System in Sichuan University. Her main research interests include the design of microwave circuits and microwave power transmission system.

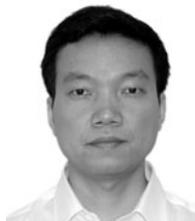

**Changjun Liu** earned his BS in Applied Physics from Hebei University of China in 1994, MS in Radio Physics in 1997, and Ph.D. in Biomedical Engineering in 2000 from Sichuan University, China. He has been a professor at the School of Electronics and Information Engineering, Sichuan University, China, since 2004. He was a postdoctoral student at Seoul National University, South Korea, from 2001 to 2002, and a visiting scholar at Ulm University, Germany, from 2006 to 2007. He was an outstanding reviewer for *IEEE Transactions on Microwave Theory and Techniques* from 2006 to 2010. Dr. Liu is focusing on microwave/radio frequency circuits, microwave wireless power transmission, and microwave power combining.